\documentclass{iau} 

\usepackage{graphicx}
\usepackage[numbers,sort,square]{natbib}
\usepackage[symbol*]{footmisc}

\title[Transitional millisecond pulsars] 
{A decade of transitional millisecond pulsars}
\vspace{-0.2 in}
\author[A.~Jaodand, J.~W.~T. Hessels, A.~Archibald]   
{Amruta Jaodand$^{1,2}$, Jason W.~T. Hessels$^{1,2}$, Anne Archibald$^2$}

\affiliation{$^1$ASTRON, the Netherlands Institute for Radio Astronomy, \\ Postbus 2, 7990 AA Dwingeloo, the Netherlands \\[\affilskip]
$^2$Astronomical Institute Anton Pannekoek, University of Amsterdam, \\ 1098XH, Amsterdam, the Netherlands}

\pubyear{2017}
\volume{337}  
\jname{Pulsar Astrophysics -­ The Next 50 Years}
\editors{P. Weltevrede, B.B.P. Perera, L. Levin Preston \& S. Sanidas, eds.}
\begin{document}

\maketitle
\vspace{-0.1 in}
\begin{abstract}
Transitional millisecond pulsars (tMSPs), which are systems that harbor a pulsar in the throes of the recycling process, have emerged as a new source class since the discovery of the first such system a decade ago. These systems switch between accretion-powered low-mass X-ray binary (LMXB) and rotation-powered radio millisecond pulsar (RMSP) states, and provide exciting avenues to understand the physical processes that spin-up neutron stars to millisecond periods. During the last decade, three tMSPs, as well as a candidate source, have been extensively probed using systematic, multi-wavelength campaigns.  Here we review the observational highlights from these campaigns and our general understanding of tMSPs. 

\end{abstract}
\vspace{-0.3 in}  
\section{Introduction}
In 2007, a 1.7-ms radio pulsar, J1023+0038 (hereafter J1023), was discovered in a binary with a low-mass companion \citep{ASR:2009}. The position was coincident with a previously identified magnetic cataclysmic variable \citep{EWB:2002,SHF:2004,TA:2005}.
When the RMSP was found, J1023 showed no evidence of an accretion disk; connecting it to the previously observed accretion state established an evolutionary link  between the LMXB and RMSP classes, as hypothesized by the pulsar recycling mechanism. According to this mechanism, a pulsar can be spun-up to millisecond periods via accretion of mass and angular momentum from a low-mass ($0.01$-$0.6$ $M_\odot$) companion undergoing Roche Lobe overflow \citep{RS:1982,ACR:1982}.

Another system IGR~J1824$-$2452 (hereafter J1824) was discovered in 2013 as an LMXB in outburst \citep{PFB:2013}, and with rotational and orbital ephemeris identified as the known RMSP J1824$-$2452I, which had apparently switched states. Within a month and a half of this LMXB outburst, J1824 reactivated as an observable radio pulsar, and established that repeated transitions between LMXB and RMSP states are possible. Soon after, in June 2013, J1023 also transitioned from RMSP to LMXB state \citep{SAB:2013,TLK:2014}. This state transition was accompanied by a factor of five brightening in $\gamma$-rays, optical brightening with observed double peaked H-$\alpha$ emission lines and disappearance of radio pulsations \citep{PAH:2014,SAH:2014}.  
Also in 2013, the {\it Fermi}-LAT source XSS~J$1227$-$4859$ (hereafter J1227), which was previously suggested to be an LMXB \citep{HSC:2011}, was shown in archival observations to have transitioned to an RMSP state in December 2012 \citep{BPH:2014,RRB:2015}. 
Together, these three systems constitute the class of well-established transitional millisecond pulsars (tMSPs). 
\vspace{-0.2 in}  
\section{Timeline of Discoveries}
Since the discovery of J1023 in 2007, tMSPs have been intensively studied using multi-wavelength campaigns \citep[e.g.,][etc.]{BAH:2011,DBF:2013,TYA:2014,JRR:2015,BAB:2015,SLN:2015, BAC:2016}.  Here, we summarize critical observations made in both the RMSP and LMXB state.  

\begin{figure}
\begin{center}
\includegraphics[trim = {0cm 3cm 0cm 2.4cm},width=\linewidth,height=0.5\linewidth]{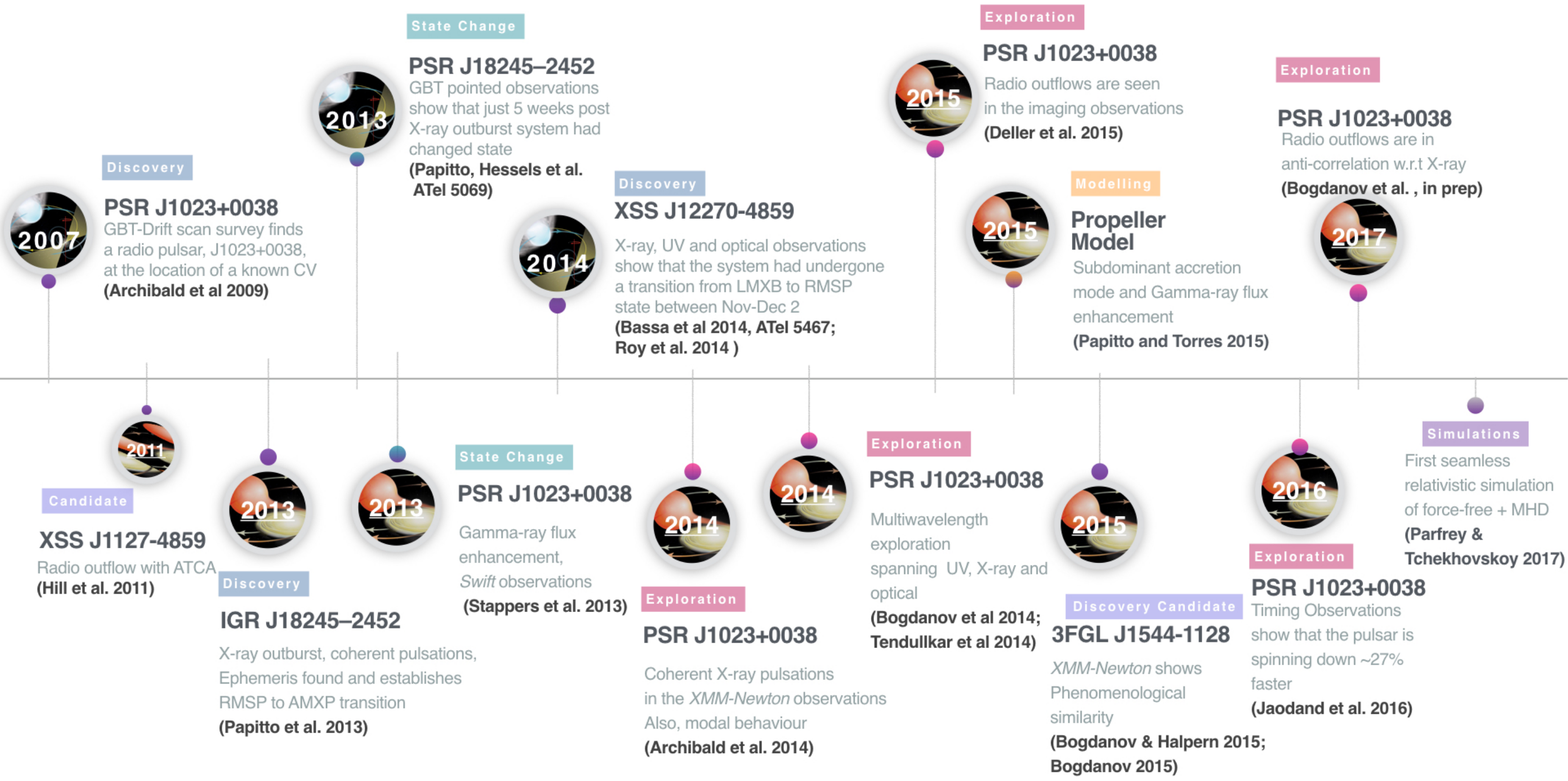}
\end{center}
\vspace{-0.075 cm}
\caption{Timeline of discoveries that significantly impacted our understanding of tMSPs. Each discovery is represented with year, publication and a short description.\vspace{-0.1 cm}}
\end{figure}
\vspace{-0.25 cm}
\subsection{tMSPs in Radio Millisecond Pulsar (RMSP) State}
Photometric observations and radio timing solutions of J1023, J1824 and J1227 have yielded companion masses of roughly 0.2, 0.3, and 0.2\,$M_\odot$ and orbital periods of 4.8, 6.9, 11.0\,hr, respectively. All three systems are ``redbacks'', i.e. millisecond pulsars with non-degenerate companions that are low-mass ($0.1-0.6$\,$M_{\odot}$), but still significantly more massive than the $<0.1$\,$M_{\odot}$, (semi)-degenerate companions of the so-called ``black widow'' pulsar binaries \citep[see][for a review of redback and black-widow pulsar binaries]{Mal:2013}.
In the RMSP state, tMSPs exhibit typical redback properties: i) orbitally modulated X-rays, possibly from an intra-binary shock \citep{AKS:2010,BAH:2011,BAA:2014}; ii) orbitally modulated optical light curve from an irradiated companion with differential temperatures on day and night side \citep{BKR:2013}; iii) radio pulsations with eclipses \citep{ASR:2009,RRB:2015,PFB:2013}; iv) $\gamma$-ray pulsations \citep{JRR:2015}; and v) non-deterministic orbital variations \citep{AKH:2013} -- these have previously been modeled as quadrupolar variation in the companion \citep{Apple:1992,AS:1994}, enhanced magnetic breaking in the companion \citep{JRP:2006}, outflows from the companion, etc. 

Timing tMSPs in the RMSP state is complicated by the eclipses and especially the non-deterministic orbital variations.  PSR J1023+0038 was timed for $\sim 4$\,yr during the RMSP state using Arecibo, GBT, WSRT, and Lovell \citep{AKH:2013}. This is currently the only tMSP with a long-term radio timing solution.  J1824 has an approximate radio timing solution, though phase connection has been challenging because it is very weak \citep{Be:2006}. J1227 has been in the RMSP state for a few years and is being monitored for timing purposes \citep{RRB:2015}. 

\vspace{-0.15 cm}
\subsection{tMSPs in Low Mass X-ray Binary (LMXB) State}
While, J1824 is the only tMSP to have gone into a luminous $L_X \gtrsim 10^{36}$\,erg~$s^{-1}$ outburst \citep{PFB:2013,Lin:2013}, tMSPs have been observed to enter a unique low-luminosity accretion state. In this LMXB state, tMSPs are fantastic probes of low-level accretion onto magnetized neutron stars. The accretion phase in tMSPs can be quasi-stable over several years. Their X-ray lightcurves can be modeled by `Low' ($L_X^{\rm Low} \sim 10^{32}$\,erg~$s^{-1}$; present for $\sim 20\%$ of the time, in the case of J1023), `High' ($L_X^{\rm High} \sim 10^{33}$\,erg~$s^{-1}$; $\sim 80\%$ for J1023) and `Flare' modes ($L_X^{\rm Flare} \sim 10^{34}$\,erg~$s^{-1}$; $< 2\%$ for J1023). Switches between different modes happen on timescales of $\lesssim 10$\,s. 

The above tMSP X-ray luminosities are usually considered as quiescent for other NS-LMXBs, but the presence of X-ray pulsations during the High mode \citep{ABP:2015,BAB:2015,JAH:2016} indicates active accretion onto the neutron star. 
PSR J1023+0038 has persisted as an LMXB since 2013 and served as an archetypal system to explore this low-level accretion regime. Radio imaging observations of tMSPs during this phase show flat-spectrum, continuum radio emission that varies on timescales of a few minutes \citep[][Jaodand et al., in prep.]{DMM:2015}. tMSPs are radio over-luminous compared to other NS-LMXBs.  Tentatively, the X-ray and radio luminosities in tMSPs are seen to be correlated such that $L_R \propto L_X^{0.7}$ in a manner similar to black holes, though more tMSPs are needed to demonstrate this correlation definitively. This suggests that tMSPs can produce radiatively inefficient outflows in a ``propeller'' mode accretion regime \citep{DMM:2015}. 
\subsubsection{X-ray Timing: PSR J1023+0038}
The persistent accretion regime, with X-ray pulsations in the High mode, enabled us to construct a precise X-ray timing solution spanning multiple years \citep[][Jaodand et al., in prep.]{JAH:2016}, which could also be compared with the radio timing solution. In the LMXB state, we could model the pulsar spin down with an enhanced constant spin-frequency derivative, such that $\dot{\nu}_{LMXB}=(1.25~\pm~0.01)\dot{\nu}_{RMSP}$. The roughly similar spin-down compared to the RMSP state hints that the pulsar wind still dominates the torque on the neutron star, with possible additional contributions from magnetospheric reconfiguration and propellering.  

\begin{figure}
\centering
\includegraphics[width=\textwidth,trim={0cm 0.25cm 0cm 0.65cm}]{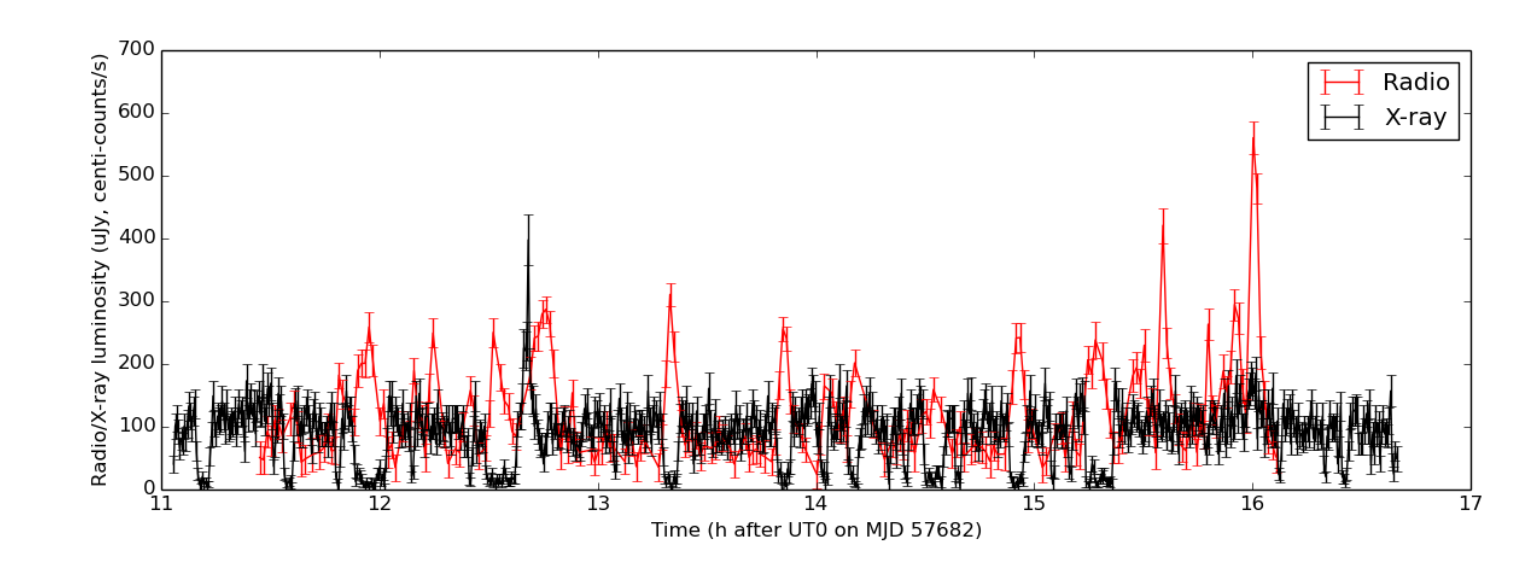}
\vfill
\caption{\footnotesize{\textit{Chandra} (black) + VLA (red) light curves of J1023 \citep{BDM:2017}. The increased radio brightness associated with each X-ray Low mode is obvious. In addition to these, the prominent X-ray flare at UT 12.7\,hr is followed by a delayed radio flare.  However, even more intense radio flaring is seen towards the end of the observation, where there is little X-ray variability.}\vspace{-0.05 cm}}
\label{fig:xr_lcorr}
\end{figure}
\vspace{-0.1 cm}
\subsubsection{Joint radio, X-ray variability: PSR~J1023+0038}
The ejecta from tMSPs should carry away angular momentum and enhance the pulsar spin-down. This in turn hints at a correlation between the X-ray mode and radio brightening. In fact, using joint \textit{Chandra} and VLA observations obtained towards late 2016, \citet{BDM:2017} have very recently seen that X-ray Low modes are coincident with radio brightening episodes, signifying that a strong radio outflow is launched (Fig. \ref{fig:xr_lcorr}). They also find that the high radio frequencies lead the low frequencies indicating an expanding, synchrotron-emitting plasma from J1023.
\vspace{-0.1 cm}
\subsubsection{Optical millisecond pulsations}
While these proceedings were in preparation, optical pulsations were discovered at J1023's rotational period \citep{APS:2017}. These pulsations were obtained by folding high resolution ($\sim$25 ns) optical data from the SiFAP20 photometer mounted atop INAF-TNG with the aforementioned X-ray timing solution. Here, the pulsed optical emission can arise from synchrotron emission by relativistic electrons and positrons i) in the pulsar magnetosphere or ii) at the site of disk and pulsar wind/magnetosphere interaction. In the future, when the other two transitional millisecond pulsars transition to LMXB state, it will be interesting to see if this behavior extends to other tMSPs. 
\vspace{-0.1 cm}
\subsection{Accretion state models}
While observational insights into tMSPs are deepening, we need new theoretical models. Recently, \cite{PT:2017} were able to simulate a full-MHD accretion disk interacting with the force-free magnetosphere of a millisecond pulsar. However, longer simulation durations are required to better understand the switching between Low and High X-ray modes in the LMXB state.  Lastly, the sustained low-luminosity accretion state and driver for switches between the LXMB and RMSP states remains a puzzle.

\centerline{}
\noindent {\small {\bf Acknowledgements} This publication has received funding from the European Union’s Horizon 2020 research and innovation programme under grant agreement No 730562 [RadioNet].}

\hrulefill
\vspace{-0.2 in}
\bibliographystyle{apj}
{\small
\bibliography{ApJ_ADJ}}

\end{document}